\documentclass[10pt,aps,prb,showpacs,superscriptaddress,twocolumn,
floatfix]{revtex4}
\usepackage{amsmath}
\usepackage{graphicx}
\begin{document}
\title{Finite band inversion of ARPES in Bi$_2$Sr$_2$CaCu$_2$O$_{8+\delta}$
 and comparison with optics}
\author{E. Schachinger}
\email{schachinger@itp.tu-graz.ac.at}
\homepage{www.itp.tu-graz.ac.at/~ewald}
\affiliation{Institute of Theoretical and Computational Physics\\
Graz University of Technology, A-8010 Graz, Austria}
\author{J.P. Carbotte}
\affiliation{Department of Physics and Astronomy, McMaster University,\\
Hamilton, Ontario, L8S 4M1 Canada}
\affiliation{Canadian Institute for Advanced Research, Toronto, Ontario,
M5G 1Z8 Canada}
\date{\today}
\begin{abstract}
Using a maximum entropy technique within a finite band Eliashberg
formalism we extract from recent high accuracy nodal direction angular
resolved photo-emission spectroscopy (ARPES) data in optimally doped
Bi$_2$Sr$_2$CaCu$_2$O$_{8+\delta}$ (Bi2212) a quasiparticle electron-boson
spectral density. Both normal and superconducting state with
$d$-wave gap symmetry are treated. Finite and infinite band results
are considered and contrasted. We compare with results obtained for
the related transport spectral density which follows from a similar
inversion of optical data. We discuss the implication of our results for
quasiparticle renormalizations in the antinodal direction.
\end{abstract}
\maketitle
\newpage
\section{Introduction}

Maximum entropy techniques have proved useful in attempts to extract
boson information either from angular resolved photo emission
spectroscopy (ARPES) data or from the optical conductivity.%
\cite{schach1,carb1,shi,zhou,hwang} What is recovered is the
electron-boson spectral density, $I^2\chi(\omega)$, which describes
the effective interaction between two electrons due to the exchange
of a boson which could be a phonon, a spin fluctuation, or some other
excitation such as a plasmon. In this way one can obtain some insight
into the nature of this interaction and by implication about
mechanism of superconductivity. In particular, coupling to a
resonance peak in $I^2\chi(\omega)$ can lead to peaks or kinks in
measured quantities. As we will see, even a structureless
background can be picked up in our inversion process. Other
approaches to the analysis of such structures have also been applied,
see for example Mishchenko and Nagaosa\cite{mish} who employed the
$t-J$-model.

ARPES
gives information on the quasiparticle self energy while optical data
can be expressed in terms of an optical self energy which is related
to, but is different from the quasiparticle self energy. There are
two main differences. The first is that optics involves the
current-current correlation function which can be expressed in terms
of a two particle Green's function and there can be vertex
corrections while ARPES requires only the knowledge
of the one-particle Green's function. Optics deals with a
transport process and transport lifetimes are not the same as
quasiparticle lifetimes. For example, the effectiveness of a scattering
process in depleting a current depends strongly on the final state,
i.e.: backward as opposed to forward scattering. Secondly, optics
deals with a momentum average while ARPES is momentum specific.
There are other
complications: ARPES measures renormalized quasiparticle energies
directly and to extract from this data the quasiparticle self energy
it is necessary to know the bare dispersion relation. Often, for
energies not too far from the chemical potential, a linear dispersion
is assumed and its slope is determined from an assumption that the
self energy crosses zero at some energy $\sim 400\,$meV in the
recent high accuracy data in Bi$_2$Sr$_2$CaCu$_2$O$_{8+\delta}$ (Bi2212).%
\cite{zhang} A similar ambiguity arises in optics: to get the
optical self energy from the reflectivity data it is necessary to
specify a value for the dielectric constant $\varepsilon$ at
infinity. Also the plasma frequency is needed and this quantity
is not so well known in the cuprates. In some determination it
requires an assumption about where the band of interest ends
as there are overlaps with higher bands, creating ambiguity.

Certainly we are dealing with a finite band situation. For a
simple first neighbor only tight binding band with hoping $t$,
the band width $W=8t$. Estimates based on tight binding fits
to local density approximation (LDA) band structure calculations
give values of the order of $350 - 450\,$meV, Ref.~[\onlinecite{mark}]
for $t$ while fits to experiment can give somewhat smaller values
of order $200\,$meV. In all cases, of course, higher nearest
neighbor hoping is also present.

So far maximum entropy inversion techniques have involved using
as an effective low energy theory for the self energy, the Eliashberg
equation generalized to include any boson-exchange mechanism and
not just phonons. But these are written for infinite bands. A
deficiency of such an approach is that the self energy cannot change
sign in an infinite band nor can its optical counterpart. On the
other hand in finite band formulations a change of sign occurs
naturally and is a robust feature of the formulation.\cite{anton,capp2}
In early ARPES experiments the real part of the quasiparticle
self energy was simply assumed to go to zero at $\sim 300\,$meV. In the
work of Meevasana {\it et al.}\cite{meev} for the
La$_{2-x}$Sr$_x$CuO$_4$ (LSCO) series the
renormalized dispersions were found to cross the bare LDA bands at
most dopings considered, with the energy of the crossing falling
roughly in the range of 400 to $600\,$meV.
Further, in the Bi2212 series of Hwang
{\it et al.}\cite{hwang1} the optical self energy was also found
to go through zero. On the theoretical side Cappelutti and
Pietronero\cite{capp2} noted that in an electron-phonon model
the real part of the self energy always goes through a zero at some
finite energy. Later Knigavko and Carbotte\cite{anton}
established that for coupling to an Einstein mode
the zero occurs at
$\sim\sqrt{\omega_E W/2}$ where $\omega_E$ is the frequency of the
oscillator. For optics the crossing is at higher energies,
$\sim\sqrt{2}$ times its quasiparticle counterpart to logarithmic
accuracy.

In this paper we start by generalizing the maximum entropy inversion
technique for the quasiparticle self energy used by Shi {\it et al.}%
\cite{shi} to finite bands. We use this new formalism to study how the
value of the band width $W$ changes results obtained for the electron-boson
spectral density $I^2\chi(\omega)$. First we consider the normal
state and generalize the procedure later on to deal with superconductivity.
For this purpose it is necessary to have finite $W$ Eliashberg
equations with $d$-wave symmetry for the superconducting gap. We
use these equations to derive from the data of Zhang {\it et al.}%
\cite{zhang} the spectral functions $I^2\chi(\omega)$ at low
temperatures in the superconducting state. The available data is for
the nodal direction only. In principle, the spectral density depends
on direction and so antinodal results would be expected to be
different. On the other hand, optics
involves a transport spectral density which is an average over
all directions in momentum space. While this means that ARPES and
optics cannot be compared directly,\cite{schach2}
we, nevertheless, use the spectral density $I^2\chi(\omega)_{\rm tr}$
taken from the optical literature to get some approximate
information on a quantity which should be close to the angular
averaged quasiparticle
self energy. We comment on the points of agreement as well as the
disagreements that are found.

In Sec.~\ref{sec:2} we provide details of our formalism and
describe results for the normal state. Section~\ref{sec:3}
deals with the superconducting state and also provides a
comparison with optics. Finally, Sec.~\ref{sec:4} gives a
brief summary and conclusions. Mathematical details are found
in Appendix~\ref{app:A}.

\section{Formalism, normal state}
\label{sec:2}

Maximum entropy techniques can be used to extract a spectral function
$I^2\chi(\omega)$ from the knowledge of the
quasiparticle self energy $\Sigma(\omega+i\delta)$ related in
integral form through a known kernel $K(\omega,\nu)$ (specified
below), namely
\begin{equation}
  \label{eq:1}
  \Sigma(\omega+i\delta) = \int\limits_{-\infty}^\infty\!
   d\nu\,K(\omega+i\delta,\nu)I^2\chi(\nu),
\end{equation}
where $I^2\chi(\omega)$ is the electron-boson spectral density which
describes the interaction of two electrons by the exchange of a boson
of energy $\nu$. The kernel $K(\omega,\nu)$ is
within Eliashberg theory in the normal state%
\cite{anton,aleks,verga,dogan,mitro,grimaldi,capp1}
\begin{eqnarray}
  \label{eq:2}
  K(\omega+i\delta,\nu) &=& \int\limits_{-\infty}^\infty\!d\omega'\,
  \frac{\tilde{N}(\omega')}{N_0(0)}\left[
  \frac{n(\nu)+f(-\omega')}{\omega-\nu-\omega'+i\delta}\right.\nonumber\\
 &&\left.
 +\frac{n(\nu)+f(\omega')}{\omega+\nu-\omega'+i\delta}\right].
\end{eqnarray}
Here, $\tilde{N}(\omega')$ is the fully renormalized electronic
density of states and carries the information on finite band
effects, $f(\omega)$ and $n(\nu)$ are the Fermi and Bose distribution
functions, and $\delta$ is an infinitesimal positive parameter.
For finite bands $\tilde{N}(\omega) \equiv N_0(0)$ and
$K(\omega+i\delta,\nu)$
reduces to a closed form. Here $N_0(0)$ is the bare electronic density
of states at the Fermi energy taken to be a constant.

 It is clear that
both real and imaginary part of $\Sigma(\omega+i\delta)$ are related to
the desired spectral function through a convolution integral to
which maximum entropy techniques apply and either can be used.
Inversion of
self energy data on $\Sigma(\omega+i\delta)$ in finite bands
requires additional information on $\tilde{N}(\omega)$. This
complication, however, can easily be handled. The quasiparticle
spectral density $A({\bf k},\omega)$ is related to the
one-particle Green's function $G({\bf k},\omega)$ by
\begin{equation}
  \label{eq:3}
  A({\bf k},\omega+i\delta) = -\frac{1}{\pi}
   \Im{\rm m}\left\{G({\bf k},\omega+i\delta)\right\},
\end{equation}
with Dyson's equation
\begin{equation}
  \label{eq:4}
  G({\bf k},\omega+i\delta) = \frac{1}{\omega+i\delta-\varepsilon_{\bf k}
   -\Sigma(\omega+i\delta)},
\end{equation}
where $\varepsilon_{\bf k}$ is the bare electron dispersion relation.
For isotropic bands we can use
$\varepsilon = \varepsilon({\bf k})$ to label the states instead of
momentum and take the simplest finite band model for the bare
density of states $N_0(\varepsilon) = 1/W$ for $\varepsilon$ in the
interval $[-W/2,W/2]$ and zero otherwise. The renormalized quasiparticle
density of states is then
\begin{eqnarray}
  \tilde{N}(\omega) &=& \int\limits_{-\infty}^\infty\!d\varepsilon\,
   N_0(\varepsilon)A(\varepsilon,\omega)\nonumber\\
  \\
  \label{eq:5}
  &=& -\frac{1}{\pi}\int\limits_{-W/2}^{W/2}\!d\varepsilon\,N_0(0)
      \Im{\rm m}\left\{\frac{1}{\omega+i\delta-\varepsilon-
      \Sigma(\omega+i\delta)}\right\}.\nonumber
\end{eqnarray}
Let us assume we know $\Sigma(\omega+i\delta)$ by some means, then
$\tilde{N}(\omega)$ is known from Eq.~\eqref{eq:5} and the kernel
Eq.~\eqref{eq:2} for the maximum entropy inversion is now definite and
the procedure can be carried out.

ARPES experiments usually provide information only on the
real part of the quasiparticle self energy, $\Sigma_1(\omega)$. Thus,
we separate Eq.~\eqref{eq:1} into its real and imaginary part and
apply the maximum entropy method to deconvolute only
\begin{equation}
  \label{eq:1a}
  \Sigma_1(\omega) = \int\limits_{-\infty}^\infty\!d\nu\,
   K(\omega,\nu)I^2\chi(\nu),
\end{equation}
with the kernel
\begin{equation}
  \label{eq:1b}
  K(\omega,\nu) = \int\limits_{-\infty}^\infty\!\!\!\!\!\!\!{\cal P}
   \,d\omega'\,\frac{\tilde{N}(\omega')}{N_0(0)}\left[
   \frac{n(\nu)-f(-\omega')}{\omega-\nu-\omega'}+
   \frac{n(\nu)+f(\omega')}{\omega+\nu-\omega'}\right].
\end{equation}
Here, $\cal P$ indicates that a principle part integral is to be
taken. Eq.~\eqref{eq:1b} differs from the one suggested by
Shi {\it et al.}\cite{shi} in their inversion work in two regards.
It contains
the fully renormalized density of states $\tilde{N}(\omega)$ which
accounts for finite band effects and the Bose distribution function
$n(\nu)$ is included because we do not want to be restricted to the
low temperature range.

If the imaginary part of the self energy, $\Sigma_2(\omega)$,
is not known from experimental data Eq.~\eqref{eq:5} cannot be applied
directly to calculate $\tilde{N}(\omega)$ and it is, therefore,
required to develop an iterative, self consistent, formalism which
will in the end allow us to extract the desired spectral function
$I^2\chi(\omega)$ in the finite band case from $\Sigma_1(\omega)$ alone.
As the kernel \eqref{eq:1b} is based on Eliashberg theory, it is only
natural to use the set of finite band $d$-wave Eliashberg equations,
\cite{jiang,choi,mars1,mars2}
as they are given in Appendix~\ref{app:A}, to calculate $\Sigma(\omega)$
for a given temperature $T$ and a given spectral function
$I^2\chi(\omega)$. When this is done, the normalized quasiparticle
density of states $\tilde{N}(\omega)$ of Eq.~\eqref{eq:5} can be
evaluated for any choice of spectral density.
The following self consistent procedure can be established:
(1) An assumption is made for $\tilde{N}(\omega)$
and the simplest one, namely  $\tilde{N}(\omega)/N_0(0)=1/W$ in the
interval $[-W/2,W/2]$ will suffice. Here, $W$ is the band width.
(2) Eq.~\eqref{eq:1a} is deconvoluted using maximum entropy techniques
and the experimental data
on $\Sigma_1(\omega)$ for a given temperature $T$. In this step
it is necessary to adjust $W$, which is an external parameter to
the deconvolution, as is $\tilde{N}(\omega)$, for best data reproduction.
The result is a first approximation to the desired spectral function
$I^2\chi(\omega)$. (3) A solution of the finite band $d$-wave Eliashberg
equations~\eqref{eq:6} based on this approximate function $I^2\chi(\omega)$,
the assumed
value of $W$ and the given temperature $T$ is generated. From this solution the
complex quasiparticle self energy $\Sigma(\omega)$ is easily calculated
and Eq.~\eqref{eq:5} can be solved to give a new fully renormalized
density of states $\tilde{N}(\omega)$ and the procedure returns to
step (1) until self consistency is reached.
Another possibility is to start with step (1) and (2) from above.
The approximate solution for $I^2\chi(\omega)$ can then be parameterized
and a least squares fit procedure using the finite band $d$-wave
Eliashberg equations~\eqref{eq:6} can be employed to get the best fit to the
experimental $\Sigma_1(\omega)$ even when the data is taken in the
superconducting state. This is our preferred method.

To get some understanding of how important finite band effects might
be in maximum entropy inversions of the quasiparticle self energy
data we proceed as follows. Hwang {\it et al.}\cite{hwang} have
obtained from optical conductivity data results for the average
transport spectral function $I^2\chi(\omega)_{\rm tr}$ in Bi2212
as a function of temperature and doping. The solid curve of
\begin{figure}[tp]
  \vspace*{-8mm}
  \includegraphics[width=9cm]{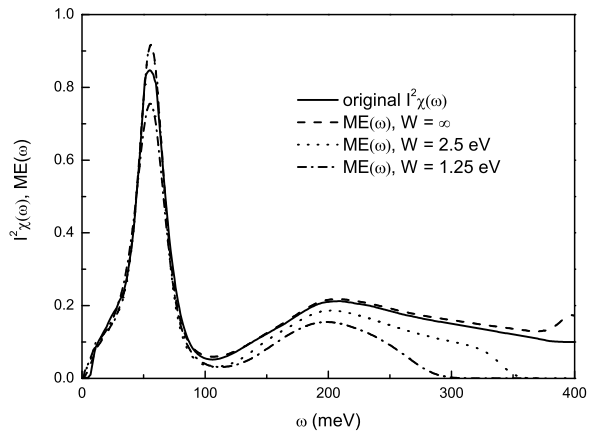}
  \caption{The electron-boson spectral density $I^2\chi(\omega)$
recovered from inversion of the real part of the quasiparticle
self energy calculated with common
$I^2\chi(\omega)$ (solid curve) but with different band width
$W = \infty$ (dashed), $W = 2.5\,$eV (dotted), and
$W=1.25\,$eV (dash-dotted). In all cases the inversion was carried
out assuming an infinite band.
}
  \label{fig:1}
\end{figure}
Fig.~\ref{fig:1} reproduces their results for an overdoped sample
($T_c = 82\,$K, labeled BI82B) at $T=26\,$K. This function is
used in Eqs.~\eqref{eq:1a} and \eqref{eq:1b} with $W\to\infty$ so
that $\tilde{N}(\omega')/N_0(0) = 1$, to obtain
the real part of the quasiparticle self energy of Eq.~\eqref{eq:1}.
[We will refer to it as the input $I^2\chi(\omega)$.]
Next, the procedure is reversed and maximum entropy techniques are used
to derive from this numerical data a new spectral function
$I^2\chi(\omega)$ which is shown as the dashed  curve in Fig.~%
\ref{fig:1} and which is seen to be almost identical to the input
curve as it must be. Next we compute again $\Sigma_1(\omega)$
from the same input $I^2\chi(\omega)$ but now we apply a finite
band cut off to the normal state version of the Eliashberg
equations~\eqref{eq:6} to generate a new set
of numerical $\Sigma_1(\omega)$ data. This new set
is again used as input for a maximum entropy
infinite band deconvolution of Eq.~\eqref{eq:1a} to yield a new model
$I^2\chi(\omega)$. Results of this procedure for $W=2.5\,$eV (dotted
curve) and $W=1.25\,$eV (dash-dotted curve) are given in Fig.~\ref{fig:1}.
It is seen that the application of a finite band cutoff to
the calculation of
$\Sigma_1(\omega)$ has a major affect at energies beyond
$100\,$meV where the new $I^2\chi(\omega)$ is considerably reduced
over its input value. This was to be expected since
inverting in an infinite band model does not account for the reduction
in self energy that is brought about by the decay in the effective
electronic density of states around the bare band edge and beyond.
It is clear that to do realistic inversions in the cuprates finite
band effects need be accounted for. This is also required if
the real part of the quasiparticle self energy is to cross zero at
some finite energy, say $400\,$meV as was assumed in the analysis of
Zhang {\it et al.}\cite{zhang} experimental data. On the other
hand, if one is mainly interested in the small energy region much
less than the zero crossing at $\sim 400\,$meV, finite band effects
make little difference except for lowering somewhat the peak around
$60\,$meV in comparison to our input spectral function.

\begin{figure}[tp]
  \vspace*{-5mm}
  \includegraphics[width=9cm]{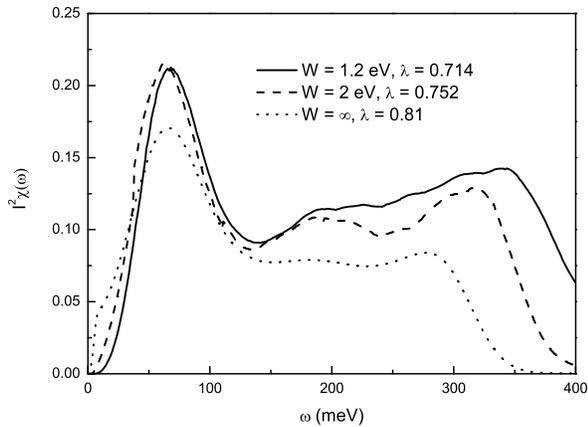}
  \caption{Result for the electron-boson spectral density
$I^2\chi(\omega)$ obtained from inversion of Zhang {\it et al.}%
\protect{\cite{zhang}}
ARPES nodal direction Bi2212 data at $T=99\,$K. The dotted curve
is for an infinite band, the green one for a band width $W=2\,$eV, and the
solid one for $W=1.2\,$eV. Note the reduction of spectral weight
beyond $\sim 100\,$meV as $W$ is increased. Values of the mass
enhancement factor are 0.81, 0.752, and 0.714, respectively.
}
  \label{fig:2}
\end{figure}

Having established our inversion technique, we next go to experiments.
We started with the $T=99\,$K data from Fig.~4a of
Ref.~[\onlinecite{zhang}] and
inverted it according to Eqs.~\eqref{eq:1a} and \eqref{eq:1b}
to recover the electron-boson spectral function $I^2\chi(\omega)$
in the specific case of the nodal direction
at $99\,$K in the normal state. Results are presented in
Fig.~\ref{fig:2}. Two different values of the band width, namely
$W = 1.2\,$eV (solid curve) and $2.0\,$eV (dashed curve) where used
together with $W=\infty$ (dotted curve).
In all cases the same experimental data appears on the left hand
side of Eq.~\eqref{eq:1a}. For the dashed and dotted curves a band
width $W=2\,$eV and $W=\infty$, respectively, was imposed on the
inversion procedure from outside. For the solid curve, the parameter $W$
\begin{figure}[tp]
  \vspace*{-5mm}
  \includegraphics[width=9cm]{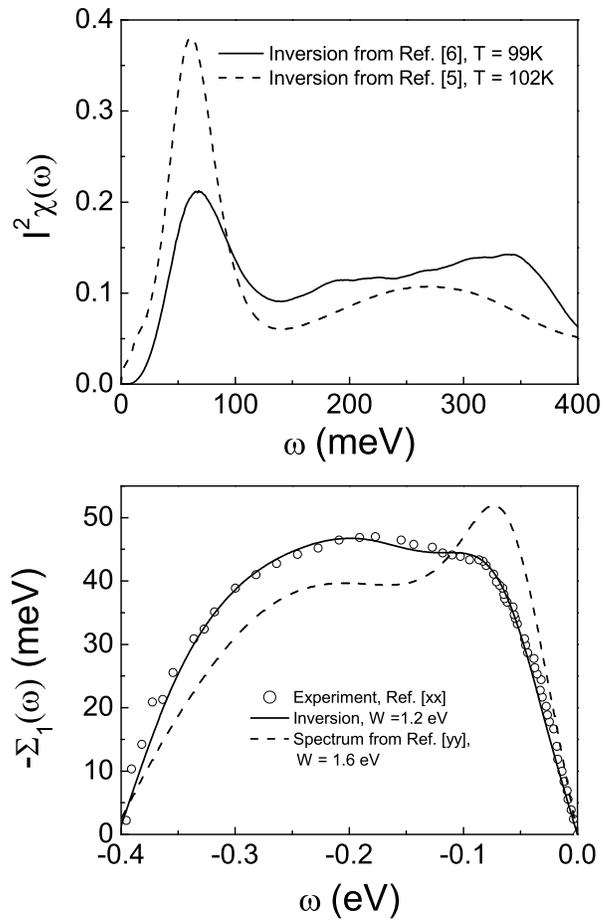}
  \caption{Top frame: Electron-boson spectral density $I^2\chi(\omega)$
from finite band inversion ($W=1.2\,$eV) of ARPES data in nodal
direction (Ref.~[\onlinecite{zhang}]) in nodal direction in the
normal state at $T=99\,$K (solid curve). The dashed curve is for
comparison and was obtained previously by the inversion of optical
data (Ref.~[\onlinecite{hwang}]) and scaled down by a factor of
0.44 so as to account for differences between quasiparticle and
transport quantities.\\
Bottom frame: Real part of the quasiparticle energy at $T=99\,$K in
the normal state. The open circles are the data of Zhang {\it et al.}%
\protect{\cite{zhang}} as read off their Fig.~4a. The solid curve
represents the result of the maximum entropy inversion of the data
assuming a band width of $W=1.2\,$eV so as to get a zero in the
self energy at $\sim 400\,$meV as in the experimental data. The dashed
curve was obtained using the $I^2\chi(\omega)$ shown by a dashed line
in the top frame. A band with of $1.6\,$eV was chosen to, again, give
a zero in $\Sigma_1(\omega)$ at $\sim 400\,$meV.
}
  \label{fig:3}
\end{figure}
was allowed to vary. Instead, a constraint, namely
that the resulting self energy $\Sigma_1(\omega)$ be zero exactly
at $\omega=400\,$meV was applied. This resulted in a value $W=1.2\,$eV.
Of course, it needs to be recognized that the ARPES experiments themselves
do not tell us where the zero in $\Sigma_1(\omega)$ occurs. Some
assumption on the bare dispersion is needed and the value $400\,$meV
while respected in our inversions for the solid curve is, therefore, model
dependent. If one had an independent, reliable estimate of the
band width then this value could be used as a constraint in the
inversion and this would yield an estimate of the energy at which
renormalized and bare band dispersions meet. Returning to
Fig.~\ref{fig:2} it should now be clear why for a fixed
set of $\Sigma_1(\omega)$ data increasing the value of $W$ leads
to smaller values of $I^2\chi(\omega)$ at higher energies.
Renormalization effects in this region can be reduced due to a
smaller value of $I^2\chi(\omega)$ which, in turn, results in a
reduced quasiparticle density of states.
Finally, we note that the mass renormalization value
$\lambda = 2\int_0^\infty d\omega\,I^2\chi(\omega)/\omega$ is of the
order 0.7 to 0.8 for optimally doped Bi2212. These values are
considerably smaller than those determined from optics as we
will discuss later. There appears to be a factor of two difference in
the magnitude between quasiparticle and transport electron-boson
spectral density.

Our best fit for $I^2\chi(\omega)$ to the data of
Ref.~[\onlinecite{zhang}] is reproduced as the solid curve in
the top frame of Fig.~\ref{fig:3} where it is compared with data
on $I^2\chi(\omega)_{\rm tr}$ (dashed curve) which was obtained
from maximum entropy inversions of optical data by
Hwang {\it et al.}\cite{hwang} for a similar optimally doped
Bi2212 sample. A scaling factor of 0.44 was applied to
$I^2\chi(\omega)_{\rm tr}$ in this case. The resulting mass enhancement
$\lambda = 1.09$ which is to be compared with the value of 0.72 we
found from ARPES. The above scaling factor was determined in
an attempt to get the best possible agreement to the ARPES
quasiparticle self energy $\Sigma_1(\omega)$ without changing
the shape of $I^2\chi(\omega)_{\rm tr}$. The corresponding self
energy is shown as a dashed line in the bottom frame of
Fig.~\ref{fig:3}. The solid curve derived from ARPES data, of
course, fits data very well within
the considered energy range of $[-0.4\,{\rm eV},0]$ for the experimental
data of Ref.~[\onlinecite{zhang}] which are indicated by open circles.
The dashed curve calculated from the rescaled
$I^2\chi(\omega)_{\rm tr}$ (dashed line in the top frame of Fig.~%
\ref{fig:3}) shows remarkable similarity. The main difference
is due to the fact that the peak in $I^2\chi(\omega)_{\rm tr}$ from
optics at $62\,$meV is stronger than the one in the ARPES data which
is at $68\,$meV. This is not unexpected since optics
produces an electron-boson spectral density which is
an average over all momenta while ARPES is momentum specific,
namely {\bf k} is in the nodal direction. If we associate the peak with
the interaction of the charge carriers with some spin fluctuations
peak around $(\pi,\pi)$ in the two-dimensional CuO Brillouin zone then
we would expect this peak to be larger for scattering in the antinodal
direction and, therefore, larger in the average function of optics
than is the nodal function of ARPES.

The good agreement found here between the ARPES and optics derived
spectral density noted in the top frame of Fig.~\ref{fig:3} shows
that both methods agree that there is strong coupling to an
excitation at $\omega=60\,$meV as well as a high energy background
which extends to $400\,$meV. This cannot be due to phonons but
finds a natural interpretation as coupling to spin fluctuations.

\section{Superconducting state and results based on optics}
\label{sec:3}

\begin{figure}[tp]
  \vspace*{-5mm}
  \includegraphics[width=9cm]{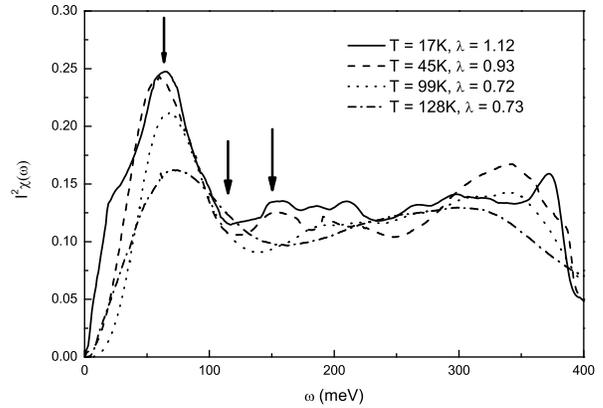}
  \caption{Results for the electron-boson spectral density
$I^2\chi(\omega)$ obtained by inversion of ARPES data by
Zhang {\it et al.}\protect{\cite{zhang}} along the nodal direction
at different temperatures, namely $T=128\,$K (dash-dotted line,
$\lambda = 0.73$), $T=99\,$K (dotted line, $\lambda = 0.72$),
$T=45\,$K (dashed line, $\lambda = 0.93$, superconducting state),
and $T=17\,$K (solid line, $\lambda=1.12$, superconducting state).
A finite band width $W=1.2\,$eV was applied.
}
  \label{fig:4}
\end{figure}

\begin{figure}[tp]
  \vspace*{-5mm}
    \includegraphics[width=9cm]{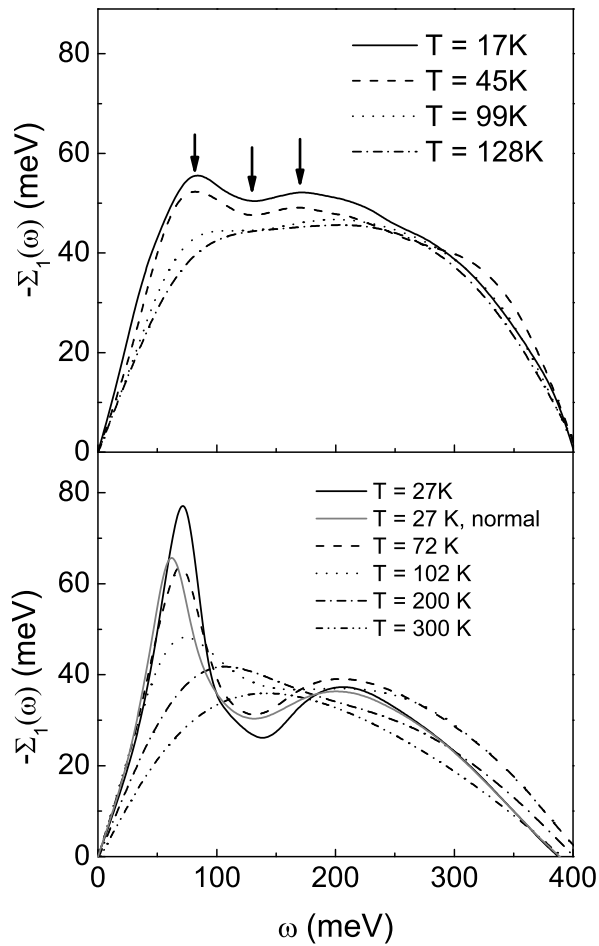}
  \caption{Top frame: The self energy $\Sigma_1(\omega)$ vs $\omega$
calculated
from the spectral densities shown in Fig.~\protect{\ref{fig:4}} which
have been found from inversion of ARPES data reported by Zhang
{\it et al.}\protect{\cite{zhang}} A finite band width of
$W=1.2\,$eV was applied.\\
Bottom frame: The self energy $\Sigma_1(\omega)$ vs $\omega$ calculated
from electron-boson spectral densities $I^2\chi(\omega)$ obtained from
optics\protect{\cite{hwang}} and scaled by a factor of 0.44. A finite
band width $W = 1.6\,$eV was applied.
}
  \label{fig:5}
\end{figure}

In the superconducting state the basic inversion procedure outlined
in the pervious section is no longer applicable because
Eqs.~\eqref{eq:1} and \eqref{eq:2} no longer apply.
We, therefore, adopted the least squares fit procedure outlined in
connection with Eqs.~\eqref{eq:1a} and \eqref{eq:1b} to perform
the inversion of the nodal direction $\Sigma_1(\omega)$ ARPES data of
Zhang {\it et al.}\cite{zhang}
 In Fig.~\ref{fig:4} we present our results of maximum entropy inversion
for temperatures
$T=128$ and $99\,$K in the normal state and 35 and $17\,$K in
the superconducting state. In all cases we get a first estimate for
$I^2\chi(\omega)$ from a deconvolution of Eq.~\eqref{eq:1a}
using the maximum entropy method. This initial form is parameterized and
then a least squares fit like procedure is applied to fit the
theoretical $\Sigma_1(\omega)$ values found from finite band
Eliashberg theory, Eqs.~\eqref{eq:6}, to experiment. The band width
is kept at $W=1.2\,$eV. All curves show
a peak around $\sim 65\,$meV which is most prominent at $17\,$K in
the superconducting state. As $T$ increases this peak broadens somewhat and
shifts towards higher energies. Besides this resonance like peak
there is a large, structured background which exists up to $400\,$meV.
It consists of a valley with its lowest point around $\sim 115\,$meV
and additional structure beginning at energies $\sim 150\,$meV.
The real part of the self energy $\Sigma_1(\omega)$ obtained from
these spectra after solution of the Eliashberg equations \eqref{eq:6}
is shown in the top frame of Fig.~\ref{fig:5}. For clarity we do not
show data but in all cases a tight fit was obtained, comparable in
quality to the fit shown in the bottom frame of
Fig.~\ref{fig:3}. As it is of considerable
interest to compare these results with optics we present the results
of additional calculations in the bottom frame of Fig.~\ref{eq:5}.
Here we used, as in the bottom frame of Fig.~\ref{fig:3}, for
$I^2\chi(\omega)$ the spectra derived from the optical data reported by
Hwang {\it et al.}\cite{hwang} for a number of temperatures, namely
$300\,$K, $200\,K$, $102\,$K, all in the normal state, and
$72\,$K and $27\,$K in the superconducting state. [The normal
state results for $T=27\,$K (solid gray line) have also been
included for comparison. In all cases a constant factor of 0.44
was used to go from transport $I^2\chi(\omega)_{\rm tr}$ to
quasiparticle $I^2\chi(\omega)$.]
A band width of $W=1.6\,$eV was chosen to
ensure a zero of $\Sigma_1(\omega)$ around $\sim 400\,$meV. The
results are very similar as to frequency and temperature
variation to those presented in the top frame of
Fig.~\ref{fig:5}. We see, again a resonance like structure at
$\sim 72\,$meV which decays with increasing temperature into a
structureless distribution for the real part of the quasiparticle self
energy $\Sigma_1(\omega)$ vs $\omega$ at $300\,$K.
It is important to note that the resonance peak is seen even above
$T_c \sim 91\,$K in both experiments.
[A comparison of the superconducting state results for $T=27\,$K
(black solid line) and the corresponding normal state results (gray
solid line) reveals that here the resonance peak is at $\sim 62\,$
meV. The gap-edge $\Delta_0$ was found to be $\sim 19.2\,$meV
and is responsible for the shift between normal and
superconducting state. For a pure $s$-wave gap $\Delta_0$ we would
expect this to be $\Delta_0$ but for $d$-wave it is less because of
the distribution in gap values.]

Finally, we note that in the optics derived case the peaks in the
lower temperature curves ($72\,$K and $27\,$K) which are in the
superconducting state are more pronounced than they are in the
nodal direction ARPES data but as we have remarked already, this
arises because optics is not momentum resolved. We expect that
the coupling to the optical resonance at $\sim 60\,$meV is larger
\begin{figure}[tp]
  \vspace*{-5mm}
      \includegraphics[width=9cm]{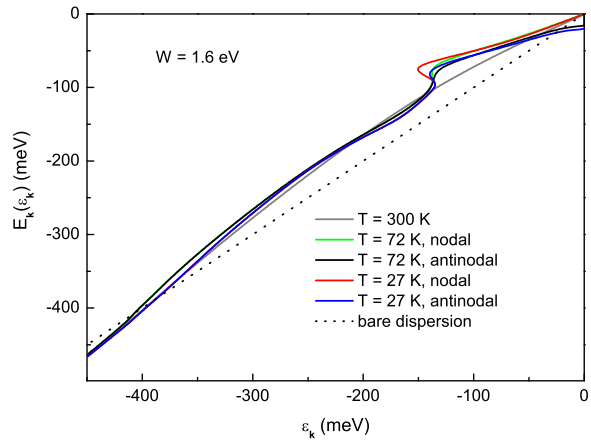}
  \caption{[Color online] The renormalized dispersion curves
$E_{\bf k}$ im meV as a function of the bare energy $\varepsilon_{\bf k}$
for various temperatures as labeled. The calculations are based
in the spectral densities $I^2\chi(\omega)_{\rm tr}$ obtained from
optics by Hwang {\it et al.}\protect{\cite{hwang}} and scaled down
by a factor of 0.44. The band with is $W=1.6\,$eV. Note that in the
superconducting state ($T=72\,$K and $T=27\,$K both, nodal and
antinodal direction are shown.
}
  \label{fig:6}
\end{figure}
in the antinodal direction and that our results are likely to be
more representative of antinodal direction ARPES data.

\begin{figure}[tp]
  \vspace*{-5mm}
        \includegraphics[width=9cm]{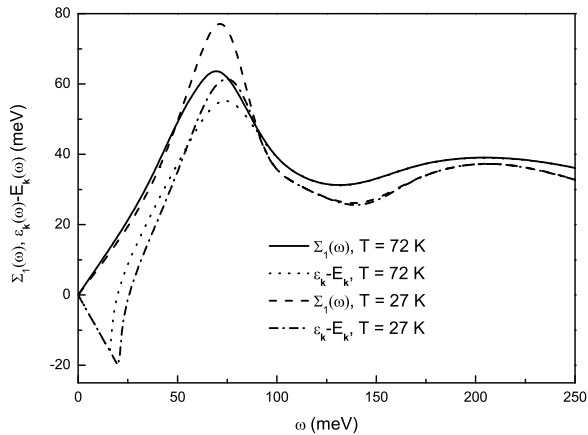}
  \caption{This illustrates the difference between the self energy
$\Sigma_1(\omega)$ and the quantity $\varepsilon_{\bf k}-E_{\bf k}(\omega)$
which are different in the superconducting state, except in the nodal
direction.
}
  \label{fig:7}
\end{figure}

In Fig.~\ref{fig:6} we compare results for the renormalized energy
$E_{\bf k}$ vs the bare energy $\varepsilon_{\bf k}$ using the
results of our Eliashberg equation solutions based on optics. With
$W=1.6\,$eV the crossing is at about $400\,$meV. For
small $\varepsilon_{\bf k}$ the renormalized energy is smaller
than its bare value which corresponds to an increase in
effective mass caused by interactions with the medium. This implies band
narrowing. By contrast beyond the zero crossing the
renormalized energy is lower than its bare value which corresponds
to band widening. For the infinite band case the quasiparticle
self energy never crosses zero and becomes small only at $\omega\to\infty$
where bare and interacting dispersion curves meet.
There is no concept of band broadening or narrowing in this case.
We wish to point out another interesting
feature of these curves. For the two low temperature curves (solid
lines and dashed lines) we are in the superconducting state and the
full Eliashberg equations \eqref{eq:6} have been solved with the
$d$-wave symmetry for the superconducting gap built in.
In our formulation, the self energy
$\Sigma(\omega+i\delta)$ is isotropic, although
its value does change from its normal state value as the gap opens.
It is
important to understand that for a superconductor, the quasiparticle
energy is given by
\begin{equation}
  \label{eq:7}
  E_{\bf k} = \sqrt{\frac{\varepsilon^2_{\bf k}+
   \tilde{\Delta}^2_1(E_{\bf k})}{Z_1^2(E_{\bf k})}},
\end{equation}
where $\tilde{\Delta}_1(E_{\bf k})$ and $Z_1(E_{\bf k})$ are the real
parts of the pairing function and of the renormalization function
$\omega Z(\omega) = \tilde{\omega}(\omega)$. Thus, at
$\varepsilon_{\bf k}=0$, $E_{\bf k} = 
\left\vert \tilde{\Delta}(E_{\bf k})/
Z_1(E_{\bf k})\right\vert$. This is zero in the nodal direction as we
can see in the two curves of Fig.~\ref{fig:6} which are labeled
`nodal'. For the antinodal direction $E_{\bf k}$ at
$\varepsilon_{\bf k} = 0$ is no longer zero. (See the two curves
in Fig.~\ref{fig:6} labeled `antinodal'.)
Finally, we note that the quasiparticle lifetime
$\Gamma(E_{\bf k})$ in Eliashberg theory is given by
\begin{equation}
  \label{eq:8}
  \Gamma(E_{\bf k}) = \frac{E_{\bf k}Z_2(E_{\bf k})}{Z_1(E_{\bf k})}
   -\frac{\tilde{\Delta}_1(E_{\bf k})\tilde{\Delta}_2(E_{\bf k})}
   {E_{\bf k}Z^2_1(E_{\bf k})}
\end{equation}
and this is not simply the imaginary part of the quasiparticle self energy
in the superconducting state which would be
$E_{\bf k}Z_2(E_{\bf k})/Z_1(E_{\bf k})$.
Because the paring function is complex
the second term in Eq.~\eqref{eq:8} is non zero and this also
contributes to the quasiparticle lifetime except in the nodal
direction for which $\tilde{\Delta}_{1,2}(E_{\bf k}) = 0$.
While Eq.~\eqref{eq:6i} determines the quasiparticle
self energy $\Sigma(\omega)$ in the normal as well as the superconducting
state, the equation $\Sigma_1(E_{bf k}) = \varepsilon_{\bf k}
-E_{\bf K}$ can only be used to determine the real part of the
quasiparticle self energy in the normal state and for the nodal
direction in the superconducting state.
This point is emphasized in Fig.~\ref{fig:7}
for the optimally doped Bi2212 sample BI96A of Hwang {\it et al.}%
\cite{hwang} with a band width
$W=1.6\,$eV. We compare in this figure $\Sigma_1(\omega)$ and
the difference $\varepsilon_{\bf k}-E_{\bf k}(\omega)$ vs $\omega$.
In the limit $\omega\to 0$ the $\Sigma_1(\omega)$ curves go to zero,
but $\varepsilon_{\bf k}-E_{\bf k}(\omega)$ does not in the antinodal
direction because there is a finite gap. This gap was already noted
in the data of Kordyuk {\it et al.}\cite{kord}
The differences are largest at low temperatures.
In all cases the two curves merge at $\omega > 100\,$meV
so that in this region the difference
$\varepsilon_{\bf k}-E_{\bf k}(\omega)$ does not represent accurately
the self energy in the superconducting state.

\section{Summary and conclusions}
\label{sec:4}

Motivated by the appearance of new high precision ARPES data on the
real part of the quasiparticle self energy in Bi2212 we describe a
procedure for extracting from this information an electron-boson
spectral density $I^2\chi(\omega)$. This work is based on Eliashberg
formalism written for a $d$-wave superconducting gap including
finite band effects. (The symbol $W$ is used for the band width.)
The superconducting state is needed because much of the data
available is for temperatures below $T_c$.
Finite bands are needed in order to get renormalized and bare
bands to cross. (They would not in an infinite band formalism.)
Furthermore, available LDA calculations as well as fits to Fermi
surfaces also indicate bands of widths of the order of some eV. In such
cases the final value of the electron-boson spectral density
$I^2\chi(\omega)$ from fits to the data is significantly affected
by $W$. While we use a maximum entropy inversion of the
convolution integral \eqref{eq:1a} to get a first numerical model
for the spectral function, in both normal and superconducting state,
the Eliashberg equations of Appendix~\ref{app:A} are employed with
a parameterized model for $I^2\chi(\omega)$ and a least squares fit to
the data.

In making comparison with spectral densities obtained from the
optical data it is necessary to recognize that optics involves
a momentum average while the ARPES is for a single momentum
direction (the nodal direction in our case). Also, the transport
$I^2\chi(\omega)_{\rm tr}$ can be different from its quasiparticle
counterpart. In making such a comparison we noted two main
differences. One, the transport spectral density is larger in
absolute values by a factor of about two. Second, the resonant peak
around $60\,$meV seen in both spectral densities is more pronounced
in optics. We believe this to reflect the importance of the
antinodal region not probed in the ARPES data of
Zhang {\it et al.}\cite{zhang} Other than these differences there
is considerable agreement between the two sets of experimental data
giving some
evidence that an Eliashberg approach can be used as a phenomenological
approach to correlate various data sets.

As is widely done in this field we interpreted the ARPES data directly
as the electron spectral function assuming the so called
`matrix element effects' described by Lindroos {\it et al.}\cite{lind}
to cause no serious distortion of the spectrum. Furthermore, large
inhomogeneities are seen in the scanning tunneling microscopy (STM)
of the Bi2212 compounds.\cite{lee} Although this technique probes
only the surface layer, the inhomogeneities could
persist in the bulk. Optics is a bulk probe and would average over the
inhomogeneities so that our analysis would reflect the average spectral
density. The good agreement with ARPES seen here could be taken as
evidence that such inhomogeneities, if important, can be treated in
an average way. An important contribution to the debate about
mechanism of superconductivity is our finding that the coupling to a high
energy background seen in optics is now confirmed in the ARPES data
by Zhang {\it et al.}\cite{zhang}

\section*{Acknowledgment}
 
Research supported in part by the Natural Sciences and Engineering
Research Council of Canada (NSERC) and by the Canadian
Institute for Advanced Research (CIAR). JPC thanks E.J. Nicol and
T. Timusk for their interest in this work.

\appendix
\section{Finite band $d$-wave Eliashberg equations}
\label{app:A}

 The generalization to $d$-wave has already been published by
Jiang {\it et al.}\cite{jiang} and has been used to describe various
aspects of the superconducting state in the cuprates. Here we include
finite bands as well. Assuming particle-hole symmetry for simplicity,
the pairs of coupled, non-linear, clean limit Eliashberg equations take
on the following form in an imaginary axis notation:\cite{choi}
\begin{subequations}
\label{eq:6}
\begin{equation}
  \label{eq:6a}
  \tilde{\omega}(i\omega_n) = \omega_n + T\sum\limits_m
   \lambda(m-n) \left\langle\frac{2\tilde{\theta}(i\omega_m,\phi')
    \tilde{\omega}(i\omega_m)}{\sqrt{\tilde{\omega}^2(i\omega_m)+
    \tilde{\Delta}^2(i\omega_m,\phi')}}\right\rangle_{\phi'}
\end{equation}
for the renormalized frequencies $\tilde{\omega}(i\omega_n)$ and
\begin{eqnarray}
  \label{eq:6b}
  \tilde{\Delta}(i\omega_n,\phi) &=& gT\sum\limits_m\cos(2\phi)
   \lambda(m-n)\nonumber\\
   \\
   &&\times
   \left\langle\frac{2\tilde{\theta}(i\omega_m,\phi')
   \tilde{\Delta}(i\omega_m,\phi')\cos(2\phi')}
   {\sqrt{\tilde{\omega}^2(i\omega_m)+\tilde{\Delta}^2(i\omega_m,
   \phi')}}\right\rangle_{\phi'}\nonumber
\end{eqnarray}
for the renormalized pairing potential $\tilde{\Delta}(i\omega_n,\phi) =
\tilde{\Delta}(i\omega_m)\cos(2\phi)$. Here, $\omega_n = \pi T (2n+1),
n = 0, \pm 1, \pm 2, ...$ are the electron Matsubara frequencies, $T$
is the temperature, $\langle ... \rangle_\phi$ denotes the average over
the polar angle $\phi$ of the two dimensional CuO Brillouin zone.
Furthermore, we have
\begin{equation}
  \label{eq:6c}
  \lambda(m-n) = 2\int\limits_0^\infty\!d\nu\,\frac{\nu I^2\chi(\nu)}
  {\nu^2+(\omega_m-\omega_n)^2}
\end{equation}
and for half-filling
\begin{equation}
  \label{eq:6d}
  \tilde{\theta}(i\omega_n,\phi) = \tan^{-1}\left[
  \frac{W}{2\sqrt{\tilde{\omega}^2(i\omega_n)+\tilde{\Delta}^2
   (i\omega_n,\phi)}}\right].
\end{equation}
We assumed here, for simplicity, the same form of $I^2\chi(\omega)$
holds for the $\tilde{\omega}$ and the $\tilde{\Delta}$ channel,
Eqs.~\eqref{eq:6a} and \eqref{eq:6b}, respectively,
and the numerical factor $g$ was introduced to account
for the fact that the projection of the general electron-boson spectral
density will in general be different in the two channels. This factor
$g$ can be determined from the linearized Eqs.~\eqref{eq:6a} and
\eqref{eq:6b} which are valid at $T=T_c$ whenever $T_c$ and
$I^2\chi(\omega)$ are known. In our particular case studied here
$g \sim 1$.

These equations are then to be analytically continued using a method
formulated by Marsiglio {\it et al.}\cite{mars1} and we get the
following result on the real $\omega$ axis:
\begin{widetext}
\begin{eqnarray}
  \tilde{\omega}(\omega) &=& \omega+iT\sum\limits_{m=0}^\infty
  \left[\lambda(\omega-i\omega_m)-\lambda(\omega+i\omega_m)\right]
  \left\langle\frac{2\tilde{\theta}(i\omega_m,\phi')
   \tilde{\omega}(i\omega_m)}{\sqrt{\tilde{\omega}^2(i\omega_m)
   +\tilde{\delta}^2(i\omega_m,\phi')}}\right\rangle_{\phi'}
  \nonumber\\
  &&+i\int\limits_{-\infty}^\infty\,dz\,I^2\chi(z)\left[n(z)-
    f(z-\omega)\right]\nonumber\\
  &&\times\left\langle\frac{2\tilde{\theta}(\omega-z)
    \tilde{\omega}(\omega-z+i\delta)}
    {\sqrt{\tilde{\omega}^2(\omega-z+i\delta)
    -\tilde{\Delta}^2(\omega-z+i\delta,\phi')}}\right\rangle_{\phi'}
  \label{eq:6e}
\end{eqnarray}
for the fully renormalized frequencies $\tilde{\omega}(\omega)$ and
\begin{eqnarray}
  \tilde{\Delta}(\omega,\phi) &=& gT\sum\limits_{m=0}^\infty
  \cos(2\phi)\left[\lambda(\omega-i\omega_m)+\lambda(\omega+i\omega_m)
  \right]\left\langle\frac{2\tilde{\theta}(i\omega_m,\phi')
  \tilde{\Delta}(i\omega_m,\phi')\cos(2\phi')}{\sqrt{
  \tilde{\omega}^2(i\omega_m)+\tilde{\Delta}^2(i\omega_m,\phi')}}
  \right\rangle_{\phi'} \nonumber\\
  &&+ig\int\limits_{-\infty}^\infty\!dz\,\cos(2\phi)I^2\chi(z)\left[
   n(z)-f(z-\omega)\right]\nonumber\\
  &&\times\left\langle\frac{2\tilde{\theta}(\omega-z)\tilde{\Delta}
   (\omega-z+i\delta,\phi')\cos(2\phi')}{\sqrt{\tilde{\omega}^2(\omega
   -z+i\delta)+\tilde{\Delta}^2(\omega-z+i\delta,\phi')}}
   \right\rangle_{\phi'}
  \label{eq:6f}
\end{eqnarray}
\end{widetext}
for the fully renormalized pairing potential $\tilde{\Delta}(\omega,\phi)$
in the real axis. Here, the function
$\tilde{\theta}(\omega,\phi)$ is defined as
\begin{eqnarray}
  \label{eq:6g}
  \tilde{\theta}(\omega,\phi) = 
   \tan^{-1}\left[\frac{iW}{2\sqrt{\tilde{\omega}^2(\omega)-
   \tilde{\Delta}^2(\omega,\phi)}}\right]
\end{eqnarray}
and
\begin{equation}
  \label{eq:6h}
  \lambda(\omega) = \int\limits_{-\infty}^\infty\!d\nu\,
  \frac{I^2\chi(\nu)}{\nu-\omega+i\delta}.
\end{equation}
\end{subequations}
Note that in cases where the square-root is complex, the branch
with positive imaginary part is to be chosen.

For the normal state only Eq.~\eqref{eq:6a} remains with
$\tilde{\Delta}(i\omega_n,\phi)\equiv 0$ which is then analytically
continued to the real axis again using Eq.~\eqref{eq:6e} with
$\tilde{\Delta}(\omega,\phi)$ set equal to zero. The quasiparticle
self energy is calculated using the relation:
\begin{equation}
  \label{eq:6i}
  \Sigma(\omega) = \omega-\tilde{\omega}(\omega)
\end{equation}
which is valid in the normal as well as in the superconducting state.


\begin{thebibliography}{99}
\bibitem{schach1}E. Schachinger, D. Neuber, and J.P. Carbotte,
\prb {\bf 73}, 184507 (2006).
\bibitem{carb1}J.P. Cabotte and E. Schachinger, Ann. Phys. {\bf 15},
585 (2006).
\bibitem{shi}Junren Shi, S.-J. Tang, Biao Wu, P.T. Sprunger,
W.I. Yang, V. Brouet, X.J. Zhou, Z. Hussain, Z.-X. Shen,
Zhenyu Zhang, and E.W. Plummer, \prl {\bf 92}, 186401 (2004).
\bibitem{zhou}X.J. Zhou, Junren Shi, T. Yoshida, T. Cuk, W.L. Yang,
V. Bruet, J. Nakamura, N. Mannella, Seiki Komiya, Yoichi Ando,
F. Zhou, W.X. Ti, J.W. Xiong, Z.X. Zhao, T. Sasagawa, T. Kakeshita,
H. Eisaki, S. Uchida, A. Fujimori, Zhenyu Zhang, E.W. Plummer,
R.B. Laughlin, Z. Hussain, and Z.-X. Shen, \prl {\bf 95},117001 (2005).
\bibitem{hwang}J. Hwang, T. Timusk, E. Schachinger, and J.P. Carbotte,
\prb {\bf 75}, 144508 (2007).
\bibitem{mish}A.S. Mishchenko and N. Nagaosa, \prl \textbf{93},
036402 (2004).
\bibitem{zhang}Wentao Zhang, Guodong Liu, Lin Zhao, Haiyun Liu,
Jianqiao Meng, Xiaoli Dong, Wei Lu, J.S. Wen, Z.J. Xu, G.G. Gu,
T. Sasagawa, Guiling Wang, Yong Zhou, Hongbo Zhang, Yong Zhou,
Xiaoyang Wang, Zhongxian Zhao, Cuangtian Chen, Zuyan Xu, and X.J. Zhou,
 arXiv:0711.1706 (unpublished).
\bibitem{mark}R.S. Markiewicz, S. Sahrakorpi, M. Lindroos, Hsin
Lin, and A. Bansil, \prb {\bf 72}, 054519 (2005).
\bibitem{anton}A. Knigavko and J.P. Carbotte, \prb {\bf 72},
035125 (2005); {\bf 73}, 125114 (2006).
\bibitem{capp2}E. Cappelluti and L. Pietronero, \prb {\bf 68},
224511 (2003).
\bibitem{meev}W. Meevasana, X.J. Zhou, S. Sahrakorpi, W.S. Lee,
W.L. Yang, K. Tanaka, N. Mannella, T. Yoshida, D.H. Lu, Y.L. Chen,
R.H. He, Hsin Lin, S. Komiya, Y. Ando, F. Zhou, W.X. Ti, J.W. Xiong,
Z.X. Zhao, T. Sasagawa, T. Kakeshita, K. Fujita, S. Uchida, H. Eisaki,
A. Fujimori, Z. Hussain, R.S. Markiewicz, A. Bansil, N. Nagaosa,
J. Zaanen, T.P. Devereaux, and Z.-X. Shen, \prb
{\bf 75}, 174506 (2007).
\bibitem{hwang1}J. Hwang, T. Timusk, and G.D. Lu, Phys. Cond.
Matter {\bf 18}, 125208 (2007); Nature (London) {\bf 427},
714 (2004).
\bibitem{schach2}E. Schachinger, J.J. Tu, and J.P. Carbotte, \prb
{\bf 67}, 214508 (2003).
\bibitem{aleks}A.S. Aleksandrov, V.N. Grebenev, and E.A. Mazur,
Pis'ma Zh. Eksp. Teor. Fiz. (JETP Lett.) \textbf{45}, 357 (1987).
\bibitem{verga}S. Verga, A. Knigavko, and F. Marsiglio, \prb
{\bf 67}, 054503 (2003).
\bibitem{dogan}F. Do\~gan and F. Marsiglio, \prb {\bf 68}, 165102 (2003).
\bibitem{mitro}B. Mitrovi\'c and J.P. Carbotte, Can. Jour. Phys. {\bf 61},
758 (1983); 789 (1983).
\bibitem{grimaldi}C. Grimaldi, E. Capelluti, and L. Pietronero,
Europhys. Lett. {\bf 42}, 667 (1998).
\bibitem{capp1}E. Cappelluti, C. Grimaldi, and L. Pietronero, \prb
{\bf 64}, 125104 (2001).
\bibitem{choi}Han-Yong Choi, \prb {\bf 53}, 8591 (1996).
\bibitem{mars1} F. Marsiglio, M. Schossmann, and J.P. Carbotte, \prb
{\bf 37}, 4965 (1988).
\bibitem{mars2}F. Marsiglio, \prb {\bf 45}, 956 (1992).
\bibitem{jiang}C. Jiang, E. Schachinger, J.P. Carbotte, D.N. Basov,
and T. Timusk, \prb {\bf 54}, 1264 (1996).
\bibitem{kord}A.A. Kordyuk, S.V. Borisenko, V.B. Zabolotnyy,
J. Geck, M. Knupfer, J. Fink, B. B\"uchner, C.T. Lin, B. Keimer,
H. Berger, A.V. Pan, Seiki Komiya, and Yoichi Ando, \prl {\bf 97},
017002 (2006), see Fig.~2a.
\bibitem{lind}M. Lindroos, S. Sahrakorpi, and A. Bansil,
\prb \textbf{65}, 054514 (2002).
\bibitem{lee}J. Lee, K. Fujita, K. McElroy, J.A. Slezak, M. Wang,
Y. Aiura, V. Bando, H. Ishikado, T. Masui, J.X. Zhu, H.V. Balatsky,
H. Eisaki, S. Uchida, and J.C. Davis Nature (London) \textbf{442}, 546
(2006).
\end{thebibliography}
\end{document}